\documentclass[conference]{IEEEtran}
\IEEEoverridecommandlockouts
\usepackage{subfiles}
\usepackage{hyperref}
\usepackage{cite}
\usepackage{amsmath,amssymb,amsfonts}
\usepackage{algorithmic}
\usepackage{graphicx}
\usepackage{textcomp}
\usepackage{xcolor}
\def\BibTeX{{\rm B\kern-.05em{\sc i\kern-.025em b}\kern-.08em
    T\kern-.1667em\lower.7ex\hbox{E}\kern-.125emX}}
\begin{document}

\title{Optimizing Cloud Deployment: Blending of IaaS and FaaS for Microservice Architecture}

\author{\IEEEauthorblockN{Nikhil Kapoor}
\IEEEauthorblockA{\textit{Electrical Engineering Department} \\
\textit{Indian Institute of Technology}\\
Delhi, India\\
nikhilkapoor259@gmail.com}
\and
\IEEEauthorblockN{Sougata Mukherjea}
\IEEEauthorblockA{\textit{Electrical Engineering Department} \\
\textit{Indian Institute of Technology}\\
Delhi, India \\
sougatam@iitd.ac.in}
}

\maketitle

\begin{abstract}
The rapid evolution of cloud computing has resulted in the adoption of hybrid deployments that blend Infrastructure-as-a-Service (IaaS) and Function-as-a-Service (FaaS) service models to optimize resource utilization, scalability, and operational efficiency. This paper presents a comprehensive study and practical implementation of a metrics-driven approach for migrating microservices from a traditional IaaS service model to a hybrid IaaS + FaaS model, using two microservice applications as case studies. The research develops an automated framework to analyze service-level performance metrics to identify microservices that are best suited for serverless execution. The findings of our research highlight the benefits and limitations of different cloud service models and provide a scalable and replicable automated methodology for optimized deployment of cloud-native applications.
\end{abstract}

\begin{IEEEkeywords}
Microservices, IaaS, FaaS, Google Cloud Platform
\end{IEEEkeywords}

\section{Introduction}
The increasing adoption of microservices-based architectures in cloud-native applications presents organizations with the challenge of optimally deploying individual services across different cloud service models, such as Infrastructure-as-a-Service (IaaS) and Function-as-a-Service (FaaS). IaaS empowers users to manage virtualized resources and orchestrate complex and stateful workloads. It provides greater control and flexibility, but demands significant management overhead \cite{Ama2016}. FaaS offers a serverless, event-driven execution model that abstracts infrastructure management and enables rapid scaling for stateless, ephemeral tasks. It offers fine-grained scalability, event-driven execution, and zero idle cost - suitable for stateless, event-driven components. However, they have limitations such as cold start latencies, stateless execution constraints, lack of control over runtime and environment, and limited support for long-running or resource-intensive tasks \cite{Sch2020}.

This challenge of optimal microservice deployment is further complicated by the complex dependencies among microservices, where the movement of a single service can have cascading effects on application performance, reliability, and cost-efficiency. In the absence of automated, data-driven approaches, organizations often rely on intuition or ad hoc decisions, which can lead to suboptimal resource utilization, increased costs, and potential violations of service-level objectives (SLO). 

Rigid adherence to a single cloud service model can lead to inefficiencies, especially in microservices-based applications where different components may have distinct operational requirements. This realization has driven interest in blended cloud service models, where applications leverage both IaaS and FaaS within a single cloud environment to maximize the strengths of each paradigm.  However, determining which microservices are best suited to switch services from IaaS to FaaS is not trivial. Manual assessment is time-consuming and error-prone, while automated solutions are still evolving. 

The primary objective of this paper is to systematically investigate and evaluate the feasibility of a metrics-driven automated approach to optimize microservice deployment from an IaaS-only service model to a blended IaaS+FaaS model, for cloud native applications.  By conducting experiments with real-world microservice-based cloud-native applications, the research seeks to provide insight into the potential, challenges, and practical considerations of adopting such an approach for modern cloud-native deployments.

The remainder of the paper is organized as follows. Section II cites related work. Section III explains the methodology of our analysis and introduces the two microservice-based applications that we use for case studies. Section IV describes the implementation of the different deployment models and Section V presents the results of our evaluation of these models. Section VI discusses a framework to automate end-to-end deployment. Finally, Section VII concludes the paper.

\section{Related Work}
The deployment of microservices on cloud platforms has been widely studied. Optimizing service placement for microservice architectures in cloud environments requires consideration of both resource demands and traffic patterns between collaborative services. \cite{Hu2019} presents algorithms for intelligent service placement that minimize resource waste while optimizing performance through strategic allocation across available infrastructure. 

Research has shown that no single resource offering can best meet all application requirements, and blending different resource offerings greatly alleviates the performance-cost optimization problem \cite{Ali2017}. Blending within IaaS has been extensively used where VMs of different sizes and transient VMs are multiplexed for cost and performance efficiency \cite{Wang2017}. Research on migrating complex stateful applications to serverless platforms has identified patterns and guidelines that facilitate placement with minimal code changes and practical performance considerations \cite{Jin2021}. Highly
parallelizable applications, such as video analytics, have been redesigned to FaaS to achieve lower latencies at reduced cost \cite{Fou2017}. 

Blending of IaaS with FaaS has also been proposed. Spock \cite{Gun2019}, SplitServe \cite{Jai2020}, and MArk \cite{Zha2019} leverage FaaS as a transition mechanism while scaling up IaaS resources and handling traffic spikes. These blended solutions have been found to be more cost-efficient compared to solutions based on a single offering type. Splice \cite{Son2022} is an automated framework for cost and performance aware blending of IaaS and FaaS services using a compiler-driven approach using annotations in the source code. Microblend \cite{Son2023} shows that blending FaaS and IaaS that takes into account microservice dependencies can significantly reduce Service Level Objective (SLO) violations and reduce costs. 

Organizations are increasingly turning to multi-cloud and hybrid cloud deployments to improve efficiency, increase flexibility, and reduce risks. However, managing such deployments presents significant challenges including operational complexity, data integration difficulties, cross-platform communication, unified monitoring, and the need for effective governance frameworks. Cross-platform deployment and orchestration engines like XFaaS \cite{Kho23} have been developed to address these challenges, enabling zero-touch deployment of functions and workflows across multiple cloud providers while using intelligent function fusion and placement logic to reduce workflow execution latency. On the other hand, \cite{Gor2024} provides a strategy for using hybrid clouds to enhance data storage and security in e-commerce applications. Infrastructure-as-code such as Terraform \cite{terraform} is being used to automate cloud deployment \cite{Tep2022}.

The widespread adoption of microservices-based architectures in cloud native systems has amplified the need for robust observability strategies to ensure system reliability and performance. Traditional monitoring methods are insufficient for cloud-native systems due to their failure points, performance bottlenecks, and resource mismanagement and require enhanced observability that focuses on collecting insights from metrics, logs, and traces \cite{cloud_native_observability}.  \cite{Fas2025} provides a comprehensive review of state-of-the-art observability frameworks and tools designed for microservice architecture.

The current literature focuses primarily on theoretical frameworks with limited empirical studies examining the practical challenges and benefits of implementing hybrid service models based on real-world performance data. This paper addresses this gap by providing a comprehensive exploration of metrics-driven service placement strategies, using actual performance data from representative microservices applications to determine deployment decisions.

\section{Methodology}
The objective of this paper is to explore metrics-driven service placement of cloud-native applications, specifically focusing on the allocation of microservices between Infrastructure-as-a-Service (IaaS) and Function-as-a-Service (FaaS) models. The approach is empirical and iterative, using real workload data and open-source monitoring tools to guide placement decisions. Our methodology used Google Online Boutique (GOB), an e-commerce platform\cite{github_gob}  and MediaReview, a movie recommendation system \cite{github_mediareview} as case study applications. These applications, with their diverse microservices and workloads, provide a solid foundation to evaluate hybrid cloud deployments. The experimental workflow involves the deployment of microservices in two configurations: IaaS-only and hybrid IaaS/FaaS to compare performance and optimize placement. Workloads that simulate realistic user interactions are applied to collect metrics that are used for service suitability assessments based on metrics such as statelessness, execution time, resource patterns, and latency sensitivity. The process is iterative, analyzing the metrics from each configuration to construct an optimized hybrid service model and validated through comparative testing. This design ensures a reproducible framework that can be applied to diverse microservices-based applications. 

\subsection{Case Study Application: Google Online Boutique}
\subsubsection{Overview}
\begin{figure}
    \centering
    \includegraphics[width=1.0\linewidth]{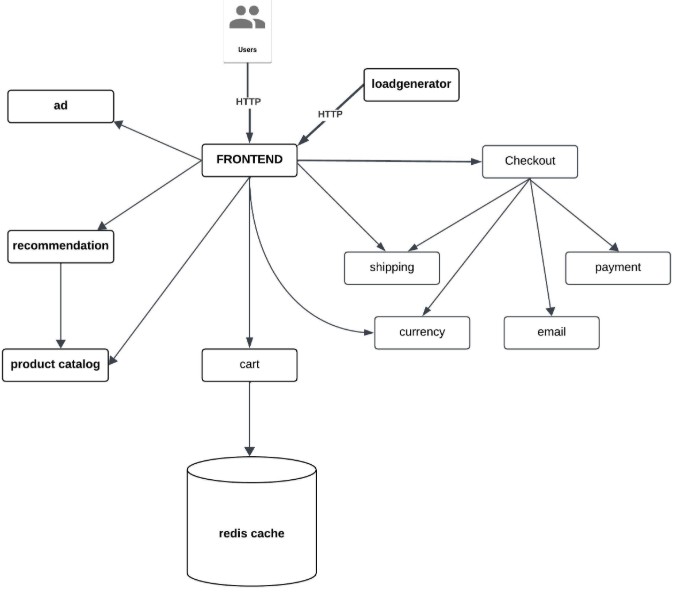}
    \caption{GOB Application standard architecture}
    \label{fig:GoB-arch}
\end{figure}

The Google Online Boutique (GOB) \cite{github_gob} is a cloud-native e-commerce application designed to demonstrate the deployment of microservices in GCP. It allows users to browse products, manage carts, and complete purchases through a web interface. The standard architecture of GOB is shown in Figure \ref{fig:GoB-arch}.

\subsubsection{Service Breakdown}
The Google Online Boutique comprises 11 microservices, each with a distinct role, implemented in various languages to ensure language-agnostic insights:
\begin{enumerate}
\item {\em frontend}: Provides the web interface for user interactions. Implemented in Go, it serves as the entry point, routing requests to back-end services. 
\item {\em product catalog}: Manages product metadata. Written in Go, it exposes gRPC endpoints, and is stateless. 
\item {\em cart}: Handles shopping cart operations, storing user carts. Uses C\# and Redis (Google Cloud Memorystore) for in-memory caching. 
\item {\em checkout}: Processes orders, coordinating with multiple services. Written in Go and  uses gRPC.
\item {\em currency}: Converts prices across currencies via external API calls. Implemented in Node.js and is stateless.
\item {\em email}: Sends order confirmation emails. Written in Python and is  stateless.
\item {\em payment}: Processes payments using external APIs. Implemented in Node.js and is stateless.
\item {\em shipping}: Calculate shipping costs. Written in Go and is stateless.
\item {\em ad}: Serves advertisements based on the user context. Implemented in Java and is stateless.
\item {\em recommendation}: Suggests products based on user history and written in Python.
\item {\em loadgenerator}: Simulates user traffic for testing and implemented in Python.
\end{enumerate}
The modularity of each service supports independent scaling and deployment.

\subsubsection{Inter-Service Communication}
Services communicate primarily via synchronous gRPC calls, with REST APIs for external interactions. The {\em frontend} service acts as the primary entry point, invoking services such as {\em product catalog, cart, checkout} via gRPC. The {\em checkout} service orchestrates a complex workflow, calling the {\em currency, email, payment}, and {\em shipping} services. The {\em recommendation} service queries {\em product catalog} service for suggestions. Cloud Pub/Sub is used for asynchronous events, such as {\em checkout service} triggering {\em email} notifications.

\subsection{Case Study Application: MediaReview}
\subsubsection{High-Level System Overview}
\begin{figure}
    \centering
    \includegraphics[width=1.0\linewidth]{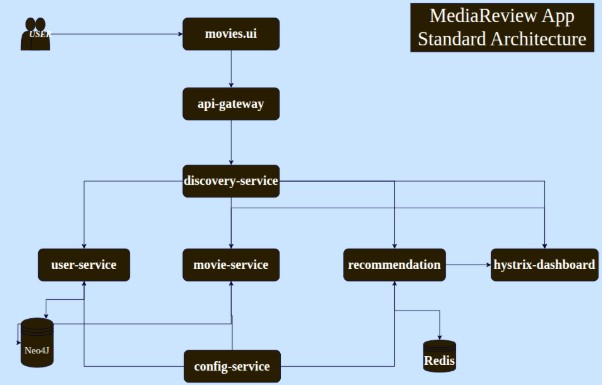}
    \caption{MediaReview application standard architecture}
    \label{fig:MediaReview_arch}
\end{figure}

MediaReview \cite{github_mediareview} is a cloud-native movie recommendation platform adapted from an open-source Spring Cloud project. It allows users to browse movies, submit ratings, and receive personalized recommendations. The architecture is shown in Figure \ref{fig:MediaReview_arch}.

\subsubsection{Service Breakdown}
MediaReview comprises eight microservices, implemented in Java using Spring Boot, including core application services and supporting infrastructure services critical to system operation.
\begin{enumerate}
\item {\em movie-service}: Manages movie metadata. Uses Spring Data Neo4j to interact with Neo4j AuraDB, exposing REST endpoints.
\item {\em user-service}: Handles user profiles and ratings. Uses Spring Data Neo4j and exposes REST endpoints.
\item {\em recommendation-service}: Generates personalized movie suggestions using Neo4j Cypher queries. 
\item {\em api-gateway}: Routes external requests to services via HTTP and uses Spring Boot.
\item  {\em consul-service}: Provides a distributed key-value store for configuration and service health monitoring and uses HashiCorp Consul.
\item {\em hystrix-dashboard}: Monitors service resilience and circuit-breaking status using Netflix Hystrix.
\item {\em discovery-service}: Manages service registration and discovery and uses Spring Cloud Netflix Eureka. 
\item {\em config-service}: Centralizes configuration management for all services and uses Spring Cloud Config. 
\end{enumerate}

\subsubsection{Inter-Service Communication}
Services communicate via synchronous REST APIs, with the {\em api-gateway} routing requests to {\em movie-service, user-service, recommendation-service, notification-service,} and the infrastructure services. The {\em recommendation-service} accesses {\em movie-service} and {\em user-service} data through Neo4j queries or REST calls. Cloud Pub/Sub facilitates asynchronous communication, with {\em recommendation-service} publishing events to trigger notifications. The {\em api-gateway} queries {\em discovery-service} (Eureka) for service endpoints and {\em config-service} for configurations. The {\em hystrix-dashboard} collects metrics from services via REST.

\section{Implementation}
This section details the practical implementation of the metrics-driven hybrid cloud service model, focusing on the deployment, monitoring, and iterative optimization of microservices across Infrastructure-as-a-Service (IaaS) and Function-as-a-Service (FaaS) platforms. The implementation followed a systematic, phased approach to assess the suitability of the service for different service models based on empirical performance data rather than intuition or ad hoc decisions. A metrics-driven framework for service placement can significantly improve both performance and cost-efficiency in microservice architectures. 

 The implementation was carried out entirely on the {\bf Google Cloud Platform (GCP)}\cite{GCP}, using {\bf Google Kubernetes Engine (GKE)} for IaaS deployment and {\bf Google Cloud Functions (GCF)} for FaaS implementation. Node.js was selected as the primary runtime for Cloud Functions due to its excellent performance characteristics for serverless workloads \cite{serverless_nodejs_advantages}.
    
We used Prometheus\cite{prometheus} to collect the metrics needed for data-driven service placement decisions. Prometheus, through the Google Managed Service for Prometheus (GMP), was configured to scrape metrics from each service. For GoB Prometheus client libraries exposed custom metrics through OpenTelemetry endpoints. For MediaReview Prometheus scraped metrics from Spring Boot Actuator endpoints.

{\bf Grafana}\cite{grafana} dashboards were created to visualize these metrics, providing a clear view of service performance and resource utilization patterns. Grafana dashboards visualized data for real-time analysis that highlight key metrics with drill-down capabilities.

To simulate realistic user traffic and collect meaningful performance metrics, {\bf Locust}\cite{locust} was used to generate the load against both applications. Locust simulated more than 50,000 requests, targeting the GOB front-end endpoints and MediaReview's recommendation-service endpoints, with steady and bursty load patterns to assess scalability.

The implementation process was structured into three distinct phases.
\begin{enumerate}
    \item IaaS-only baseline deployment and metrics collection
    \item Metrics analysis and initial service placement in FaaS
    \item Hybrid model evaluation and iterative optimization. 
\end{enumerate}

\subsection{Phase 1: IaaS-Only Deployment}
\subsubsection{Containerization and Kubernetes Deployment}
All microservices were initially deployed on GKE. 
\begin{itemize}
\item \textbf{Google Online Boutique}: All 11 services were containerized using Docker and run on GKE as Kubernetes pods. Docker images were built for each service using tailored Dockerfiles. Existing Dockerfiles supported multistage builds (for example, Go for {\em frontend}, Node.js for {\em currency}). Stateful services like {\em cart} service connect to Google Cloud Memorystore (Redis) for caching, while others operate statelessly. Service objects provide intra-cluster communication via DNS. GKE ConfigMaps store service configurations (for example, API endpoints, Redis credentials), ensuring consistency across pods. GKE’s native Kubernetes Service objects and DNS handle service discovery.
\item \textbf{MediaReview}: All 8 services run on GKE as Kubernetes pods.  Custom Dockerfiles were created for Spring Boot services (for example, {\em movie-service, notification-service}), defining Java 17, Maven dependencies, and JAR execution. The images were pushed to the Google Container Registry (GCR) using the docker push command. Stateful services such as {\em movie-service}, and {\em user-service} connect to Neo4j AuraDB for graph data and Redis for query caching. Infrastructure services ({\em consul-service, discovery-service, config-servic}e) support configuration and discovery.  The {\em config-service}, backed by {\em consul-service}, centralizes service configurations, supplemented by GKE ConfigMaps for Kubernetes-specific settings (e.g., Neo4j credentials). The {\em discovery-service} (Eureka) and {\em consul-service} manage service registration, with GKE’s Kubernetes Service objects providing native discovery for GKE-deployed services. 
\end{itemize}
Kubernetes manifests defined deployments and services for each service.  Services exposed ports (for example, $8080$ for {\em movie-service} of MediaReview and  $3550$ for GoB's {\em frontend}).

\subsubsection{Metrics Collection}
Prometheus, through the Google Managed Service for Prometheus (GMP), scraped endpoints every 15 seconds.  Prometheus was configured to scrape metrics from all services, with a focus on key metrics like
request latency, CPU and memory utilization, as well as error rates and throughput.

\subsection{Phase 2: Metrics Analysis and Service Placement}
\subsubsection{Metrics-Driven Service Selection}
\cite{Mal2020} shows that GCP offers a 14-40\% cost advantage for equivalent performance levels when comparing FaaS to IaaS for certain workloads. Therefore, to explore the suitability of FaaS for our applications after collecting baseline metrics from the IaaS-only deployment, a systematic analysis was performed to identify services suitable for placement on FaaS.  The following criteria were used to assess service suitability:
\begin{enumerate}
    \item Statelessness: Services that require a persistent state in memory were identified as not suitable for FaaS.
    \item Execution Duration: Services with execution times consistently within the GCF timeout limit (540s) were considered \cite{nodejs_runtime}.
    \item Resource Utilization Patterns: Services with sporadic or bursty workloads were identified as better candidates for FaaS.
    \item Latency Sensitivity: Services critical to user experience were evaluated for impact from cold start.
\end{enumerate}
Based on this analysis, the following services were selected for placement in GCF:
\begin{itemize}
\item  Google Online Boutique:
\begin{itemize}
    \item {\em email}: Stateless, event-driven, non-latency-critical
    \item {\em currency}: Stateless, low CPU usage, sporadic traffic
    \item {\em ad}: Bursty traffic pattern, stateless operation.
    \item {\em shipping}: independent, stateless, event-driven.
\end{itemize}
\item MediaReview:
\begin{itemize}
    \item {\em recommendation-service}: Event-driven, compute-intensive, but within the GCF limits.
    \item {\em user-service}: Event-driven, stateless.
\end{itemize}
\end{itemize}
Services with complex synchronous dependencies (for example, {\em checkout}, {\em api-gateway}) stayed on GKE for reliability. This analysis guided the hybrid configuration, optimizing placement across both applications. 

\subsubsection{Service Adaptation for FaaS}
Selected services were adapted for GCF deployment, following the best practices outlined in the Google Cloud documentation \cite{nodejs_runtime}. 
\begin{enumerate}
    \item Code Refactoring: Services were refactored to comply with GCF's programming model.
    \item Externalization of state: Any state was moved to external services (e.g., Firestore, Redis).
    \item Trigger Configuration: HTTP and event triggers were configured based on the service requirements.
\end{enumerate}

\subsection{Phase 3: Hybrid Model Evaluation and Optimization}
\subsubsection{Hybrid Deployment Configuration}
After migrating selected services to GCF, a hybrid service model was developed with services running on both GKE and GCF. This approach is supported by research showing that blending IaaS and FaaS that takes into account microservice dependencies can significantly improve performance, reduce Service Level Objective (SLO) violations, and reduce costs \cite{Son2023}.

The hybrid configuration allocated services based on metrics analysis.
\begin{figure}
    \centering
    \includegraphics[width=1.0\linewidth]{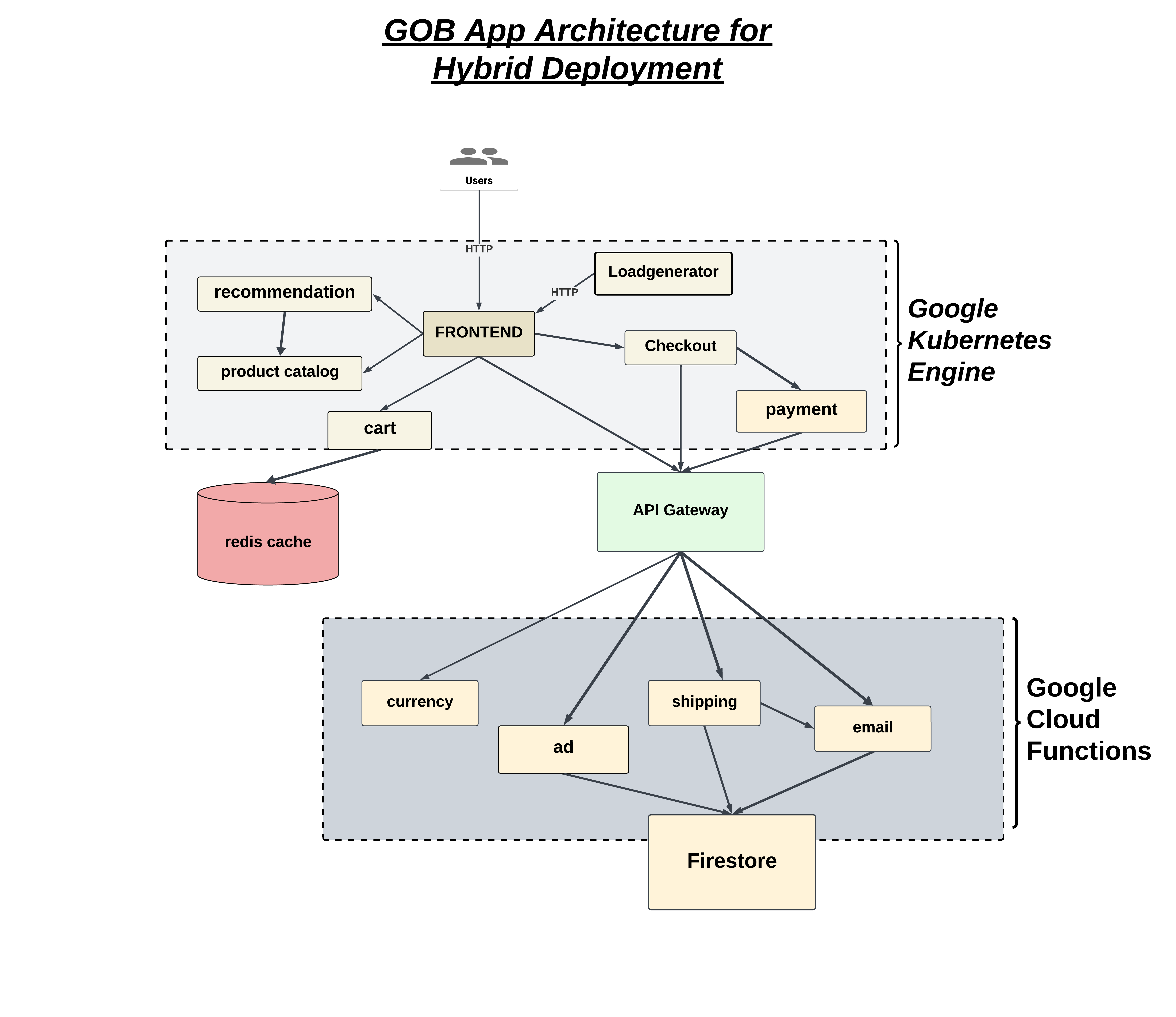}
    \caption{Hybrid IaaS/FaaS Architecture of GOB}
    \label{fig:HybridArchGOB}
\end{figure}
\begin{itemize}
\item \textbf{Google Online Boutique}: 
\begin{itemize}
\item \textbf{GKE}: The stateful services, namely {\em front-end, cart, checkout, product catalog, payment, shipping,} and {\em recommendation}, were run on GKE.
\item \textbf{GCF}: Event-driven services {\em email, currency, shipping} and {\em ad} ran on GCF. 
\end{itemize}
The hybrid architecture for Google Online Boutique is shown in Figure \ref{fig:HybridArchGOB}
\begin{figure}
    \centering
    \includegraphics[width=1\linewidth]{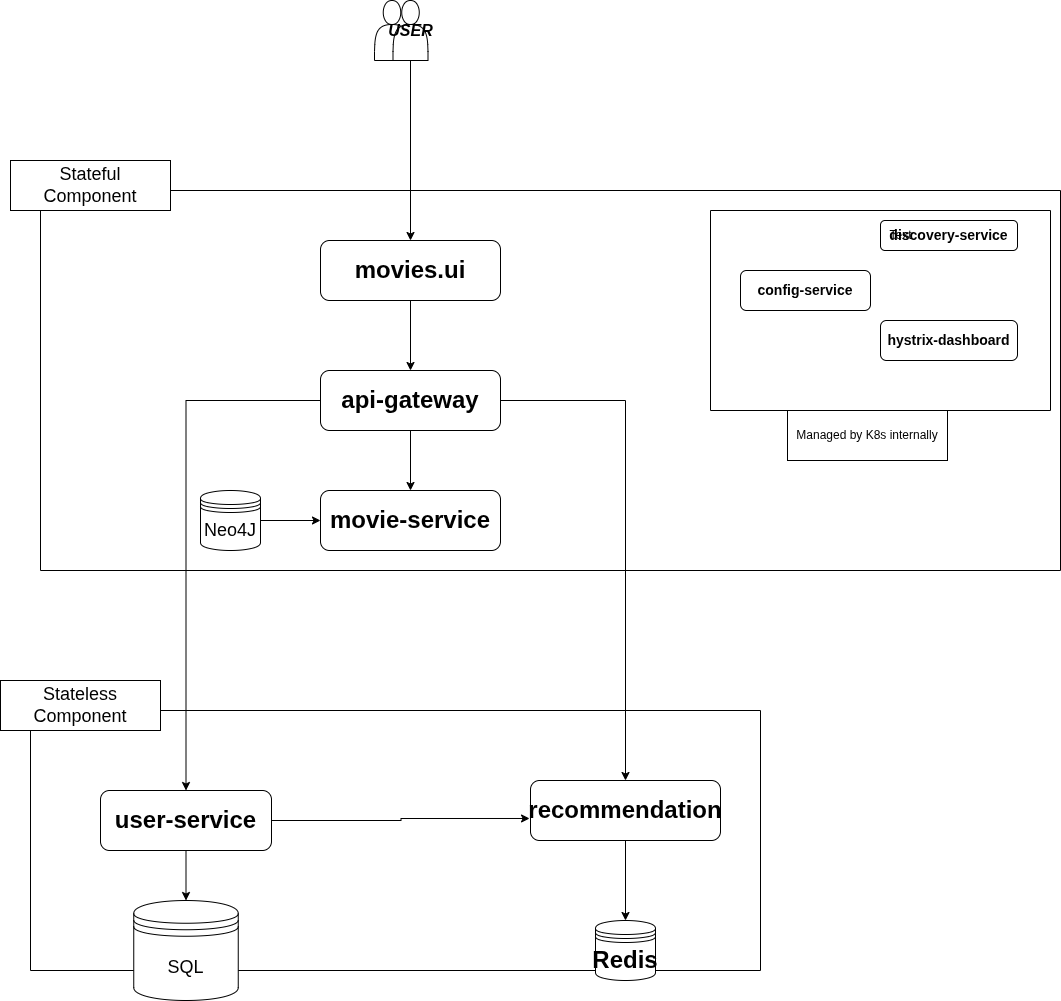}
    \caption{Hybrid IaaS/FaaS Architecture of MediaReview}
    \label{fig:HybridArchMR}
\end{figure}
\item \textbf{MediaReview}: 
\begin{itemize}
\item \textbf{GKE}: Stateful and infrastructure services, namely {\em movie-service, user-service, api-gateway,} and  {\em consul-service} ran on GKE.
\item \textbf{GCF}: Event-driven service, {\em recommendation-service} and {\em notification-service} ran on GCF.
\end{itemize}
The hybrid architecture of MediaReview is shown in Figure \ref{fig:HybridArchMR}.
\end{itemize}

\subsubsection{Communication between IaaS and FaaS}
Enabling seamless communication between services running on GKE and GCF was a critical aspect of the hybrid architecture. Two primary integration patterns were implemented:
\begin{enumerate}
    \item Synchronous communication: For request-response patterns, HTTP endpoints with Serverless VPC access were used for secure communication. Rule-based orchestration, coded in {\em checkout} (Go) and {\em api-gateway} (Java), optimizes inter-service interactions. 
    \item Asynchronous Communication: For event-driven interactions, Google Cloud Pub/Sub was used as a message broker. 
    Cloud Pub/Sub enabled asynchronous communication (for example, GOB’s {\em checkout} to {\em email} and  MediaReview’s {\em recommendation-service} to {\em notification-service}).
\end{enumerate}
This integration approach is supported by research from Google Cloud, which recommends using VPC connectors and Pub/Sub for secure and reliable communication between GKE and GCF  \cite{gke_enterprise_overview}.

\subsubsection{Metrics Collection and Analysis}
The same monitoring infrastructure used in Phase 1 was extended to collect metrics from both the GKE and GCF services. Based on specifications from \cite{Ust2023}, key metrics collected from GCF services include execution time, memory usage, error rates, as well as cold start frequency and duration. 

\subsubsection{Cold Start Latency}
In Function-as-a-Service, a major problem is cold start latency, which refers to the extra time it takes to handle a request when a function is invoked for the first time (or after being idle). GCF services experienced significant cold start latency, particularly for infrequent requests, with delays of up to 5 seconds observed.  Various approaches have been proposed to reduce this problem \cite{Gol2024}. To reduce cold start latency we have implemented minimum instances for critical services and optimized function code for faster initialization.  It has been shown that startup CPU boost can reduce cold start times by up to 30\% for Node.js functions \cite{cpu_boost_GCF}.

\section{Evaluation}
We evaluated the performance of IaaS only and hybrid deployments for the two microservice-based applications. The results demonstrate the effectiveness of the proposed approach in optimizing microservice deployment across Google Kubernetes Engine (GKE) and Google Cloud Functions (GCF) platforms through systematic performance analysis.

All experiments used Prometheus monitoring and Grafana visualization tools to collect comprehensive performance data in the Google Online Boutique and MediaReview applications. The evaluation methodology used Locust for load generation, enabling precise control over traffic patterns essential for validating hybrid deployment strategies. Load testing was performed for 200 concurrent users for 25 minutes, conducted in two load patterns: linear and bursty.  The pattern represents different real-world usage scenarios that modern microservices applications typically encounter.

\subsection{Comparative Performance Analysis: IaaS vs Hybrid Deployment}
The comprehensive evaluation of both case study applications across IaaS-only and hybrid deployment models demonstrates significant performance improvements achieved through metrics-driven service placement strategies. The empirical results provide compelling evidence for the effectiveness of the proposed hybrid architecture approach.

\begin{itemize}
\item {\bf Google Online Boutique:}
The four services migrated from GKE to GCF in the Google Online Boutique application showed consistent latency improvements across all load patterns:
\begin{itemize}
    \item {\em email}: 21.2\% latency improvement (495ms → 390ms)
    \item {\em currency}: 14.5\% latency improvement (620ms → 530ms)
    \item {\em ad}: 17.2\% latency improvement (580ms → 480ms)
    \item {\em shipping}: 9.6\% latency improvement (650ms → 587ms)
\end{itemize}
\item{\bf MediaReview:}
The services migrated from GKE to GCF in MediaReview application also demonstrated similar performance improvements:
\begin{itemize}
    \item{\em user-service}: 5.5\% latency improvement (720ms → 680ms)
    \item {\em recommendation-service}: 14.8\% latency improvement (880ms → 750ms)
\end{itemize}
\end{itemize}
These improvements directly correlate with the services' baseline characteristics, particularly their low resource utilization and stateless operation. Consistent performance gains across different services validate the selection criteria established in the methodology phase. The consistency of these improvements across different application domains provides strong empirical validation for the metrics-driven service placement approach. Services with similar characteristics (low resource utilization, statelessness, sub-500ms baseline latency) achieved comparable performance gains regardless of their specific functionality or application context.

\subsection{Load Pattern Performance Analysis}
The load testing approach revealed important information on optimal service placement under different traffic conditions. Table \ref{tab:load_testing} shows the performance in load testing.

\begin{table*}[htbp]
\centering
\caption{Latency and Performance Metrics of the application in load testing}
\label{tab:load_testing}
\resizebox{\textwidth}{!}{%
\begin{tabular}{|l|l|c|c|c|c|c|c|}
\hline
\textbf{Application} & \textbf{Deployment} & \textbf{\# Load} & \textbf{\# Requests} & \textbf{\# Fails} & \textbf{Median latency (ms)} & \textbf{Average latency (ms)} \\
\hline
{GoB} & {IaaS only} & {Linear} & {53371} & {12} & {990}  & {1093.41} \\
{GoB} & {Hybrid}    & {Linear} & {51847} & {8} & {820}  & {1034.23} \\
{GoB} & {Hybrid}    & {Bursty} & {53100} & {34} & {890}  & {986.23} \\
{MediaReview} & {IaaS only} & {Linear} & {53206} & {47} & {710}  & {816.42} \\
{MediaReview} & {Hybrid}    & {Linear} & {52200} & {24} & {620}  & {711.23} \\
{MediaReview} & {Hybrid}    & {Bursty} & {54900} & {42} & {680}  & {780.90} \\
\hline
\end{tabular}%
}
\end{table*}

\subsubsection{Linear Load Pattern Performance}
 Linear load testing demonstrated consistent performance improvements across all migrated services in both applications. The gradual increase in traffic allowed for precise measurement of improvements in service behavior after placement, and all FaaS-migrated services showed 10- 20\% latency reductions compared to their IaaS counterparts.

\subsubsection{Bursty Load Pattern Performance}
Bursty traffic patterns revealed the greatest advantages for FaaS-migrated services. All services in GCP are auto-scaled during bursty load, providing superior resource efficiency and performance optimization. The results show that even for bursty load the hybrid deployment latency is lower than IaaS only deployment for linear load. In addition, the reduction of resource waste during idle periods demonstrates the significant economic benefits of the hybrid approach for services with irregular traffic patterns.

\section{Automation}
 Although the manual application of our methodology proved to be effective in controlled experiments, the manual transition of microservices is labor intensive and error-prone, particularly in environments characterized by a large number of services and frequent deployment cycles. A framework that automates the placement workflow, which includes metrics collection, suitability analysis, automated deployment, and dynamic service reconfiguration, is essential to ensure consistency, repeatability, and scalability in the transition process. 
 
\subsection{Methodology}
\begin{figure}
    \centering
    \includegraphics[width=1\linewidth]{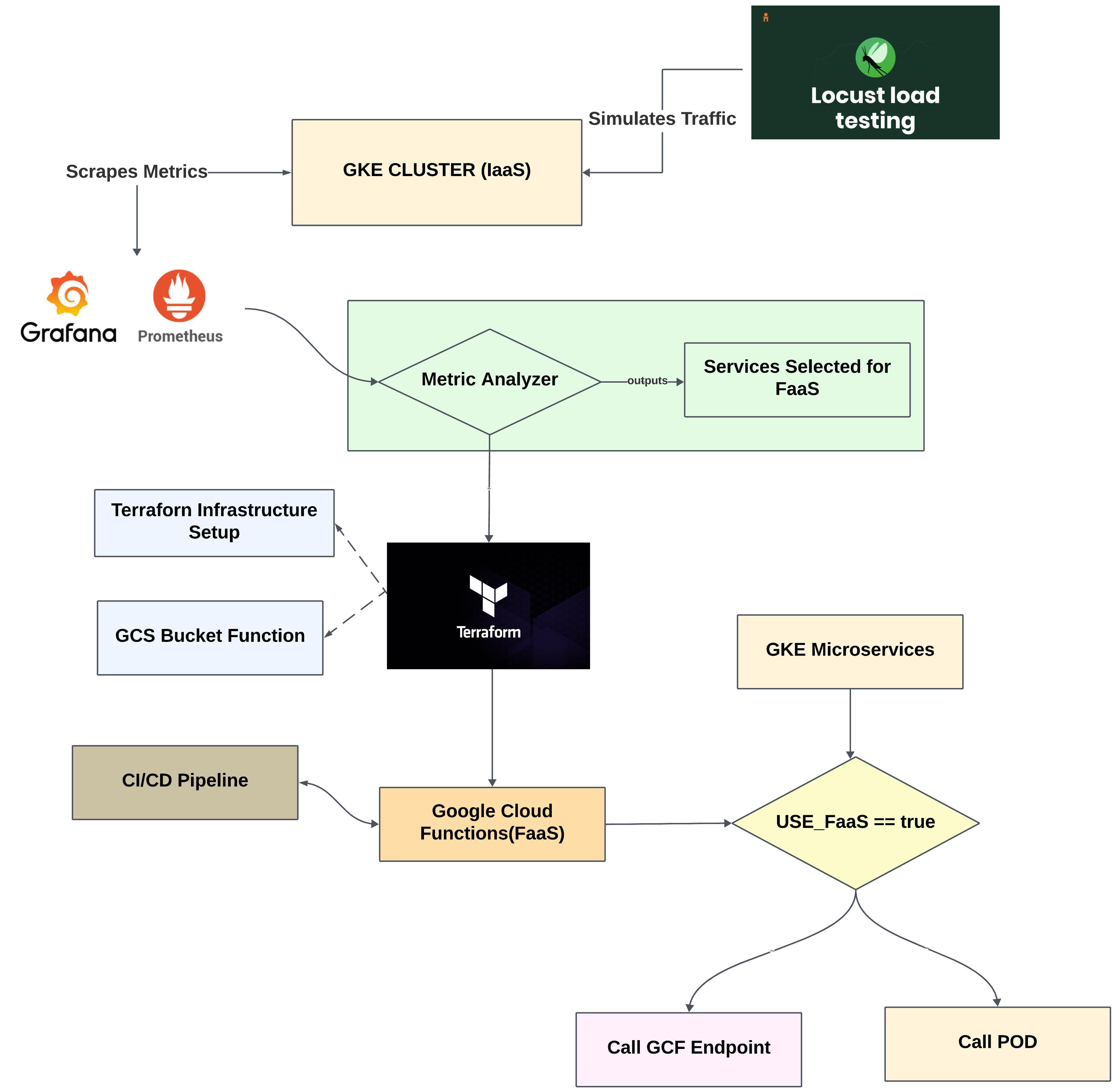}
    \caption{Automated Framework for Hybrid Deployment}
    \label{fig:automation}
\end{figure}

The automation framework shown in Figure~\ref{fig:automation}, is configured as a sequence of interconnected modules, each responsible for a distinct phase of the deployment pipeline. The process begins with the generation of synthetic traffic using Locust, which ensures that the metrics collected are representative of real-world usage patterns. Prometheus, integrated with Grafana, is used to extract key performance metrics from the Kubernetes cluster at regular intervals. The metrics of interest include CPU and memory usage, response latency (with particular attention to the 99th percentile), error rates, and request counts. These metrics are aggregated over a sliding window of {\em 10 minutes}, with data points collected every {\em 30 seconds}, providing a complete view of the behavior of the service.

The suitability of each service for transition to a FaaS environment is determined by a rule-based analysis implemented as a Bash script. The placement criteria are determined by the empirical findings discussed previously. Services that are stateless and event-driven, with average CPU and memory usage less than $30\%$ and response latency less than $500$ milliseconds, are prioritized for transition. However, the framework allows flexibility in these thresholds, accommodating services with slightly higher resource usage or latency if other factors are favorable. {\em Error rates} are also considered, with target threshold under $0.1\%$ for user-facing services, based on industry best practices.

Once a set of candidate services has been identified, the framework moves on to the packaging and deployment phase. The source code for each service is extracted from the version control repository, the dependencies are resolved, and the code is packaged into a ZIP archive suitable for deployment in Google Cloud Functions. The deployment process is orchestrated using Terraform, with environment variables such as the service name and handler function passed dynamically to the configuration scripts. The packaged code is first uploaded to Google Cloud Storage, from where it is deployed to Google Cloud Functions. Upon successful deployment, the public URL of the newly created function is retrieved and injected into the relevant microservices as an environment variable, enabling dynamic routing to the FaaS endpoint.

The entire workflow is integrated with a CI/CD pipeline implemented using GitHub Actions \cite{GithubActions}. This integration enables automated execution of the placement process in response to code changes or on a scheduled basis. Although explicit rollback mechanisms are not currently implemented, the framework continuously monitors service metrics and can revert routing to the original Kubernetes pod if anomalies are detected.

\begin{table}[h]
\centering
\caption{Comparison of Manual and Automated Hybrid Deployment Processes}
\label{tab:automation_comparison}
\begin{tabular}{|l|c|c|}
\hline
\textbf{Metric} & \textbf{Manual} & \textbf{Automated} \\
\hline
Average Placement Time (min)& 20 & 6 \\\hline
Configuration Errors (\%) & 8& 2\\\hline
Average Rollback Time(min) & 25 & 5 \\ \hline
\end{tabular}
\end{table}

\subsection{Validation}
The framework was validated using the Google Online Boutique application, demonstrating its effectiveness in reducing manual effort and improving placement reliability. The framework was evaluated over multiple cycles, with metrics collected during transition time, configuration errors, rollback duration, and the number of services migrated per hour. The results, summarized in Table~\ref{tab:automation_comparison}, indicate a substantial reduction in manual effort and deployment time, as well as improved consistency and reliability. Automation resulted in a marked decrease in placement time and configuration errors while enabling repeatable and scalable transitions to FaaS.

The automation framework utilizes a unified pipeline to ensure that migration decisions are data driven, deployments are consistent, and service disruptions are minimized. The modular design of the framework allows for extension to additional cloud providers and the incorporation of more sophisticated analysis techniques in the future. 

\section{Conclusion}
This paper investigates the hybrid deployment of microservices combining Infrastructure-as-a-Service (IaaS) and Function-as-a-Service (FaaS) platforms. The core problem addressed is whether a systematic, metrics-driven service placement approach can effectively determine the suitability of services for FaaS placement, thus improving performance and scalability. Our methodology presents a reproducible, automated framework for metrics-driven service placement in hybrid IaaS/FaaS cloud environments, demonstrated through GOB and MediaReview case studies. We determined metrics like statelessness and latency for multiple microservices across IaaS-only, FaaS-only, and hybrid configurations, using open-source tools (Prometheus, Grafana, Locust) to determine the optimum  placement. The results confirm that services with low resource consumption, statelessness, and bursty or sporadic traffic patterns are suitable for FaaS, and that strategic placement based on empirical metrics can significantly enhance system response times and resource efficiency.  Our research documents the practical challenges, architectural considerations, and optimization strategies required for successful hybrid deployment and provides a practical, automated framework for organizations seeking to optimize their cloud-native applications. 

Based on our findings, several avenues for future research and development are evident.
\begin{itemize}
    \item Develop adaptive algorithms that continuously monitor performance metrics and traffic patterns to reconfigure service placement in real time.
    \item Conduct longitudinal studies to assess the operational and economic impacts of hybrid deployment strategies over extended periods, considering evolving workloads and cloud pricing models.
    \item Explore predictive analytics for workload forecasting and proactive resource provisioning using AI/ML techniques.
\end{itemize}

\bibliographystyle{IEEEtran}
\bibliography{IEEEabrv,refs}

\begin{thebibliography}{10}
\providecommand{\url}[1]{#1}
\csname url@samestyle\endcsname
\providecommand{\newblock}{\relax}
\providecommand{\bibinfo}[2]{#2}
\providecommand{\BIBentrySTDinterwordspacing}{\spaceskip=0pt\relax}
\providecommand{\BIBentryALTinterwordstretchfactor}{4}
\providecommand{\BIBentryALTinterwordspacing}{\spaceskip=\fontdimen2\font plus
\BIBentryALTinterwordstretchfactor\fontdimen3\font minus \fontdimen4\font\relax}
\providecommand{\BIBforeignlanguage}[2]{{%
\expandafter\ifx\csname l@#1\endcsname\relax
\typeout{** WARNING: IEEEtran.bst: No hyphenation pattern has been}%
\typeout{** loaded for the language `#1'. Using the pattern for}%
\typeout{** the default language instead.}%
\else
\language=\csname l@#1\endcsname
\fi
#2}}
\providecommand{\BIBdecl}{\relax}
\BIBdecl

\bibitem{Ama2016}
M.~Amaral, J.~Polo, D.~Carrera, I.~Mohomed, M.~Unuvar, and M.~Steinder, ``{Performance Evaluation of Microservices Architectures using Containers},'' in \emph{Proceedings of the IEEE 14th International Symposium on Network Computing and Applications (NCA)}, 2016.

\bibitem{Sch2020}
J.~Scheuner and P.~Leitner, ``{Function-as-a-Service Performance Evaluation: A Multivocal Literature Review},'' \emph{Journal of Systems and Software}, vol. 170, p. 110708, 2020.

\bibitem{Hu2019}
Y.~Hu, C.~de~Laat, and Z.~Zhao, ``{Optimizing Service Placement for Microservice Architecture in Clouds},'' \emph{Applied Sciences}, vol.~9, no.~21, 2019.

\bibitem{Ali2017}
O.~Alipourfard, H.~Liu, J.~Chen, S.~Venkataraman, M.~Yu, and M.~Zhang, ``{Cherrypick: Adaptively unearthing the best Cloud configurations for Big Data Analytics},'' in \emph{Proceedings of the Proceedings of the 14th USENIX Conference on Networked Systems Design and Implementation (NSDI'17)}, 2017.

\bibitem{Wang2017}
C.~Wang, B.~Urgaonkar, A.~Gupta, G.~Kesidis, and Q.~Liang, ``{Exploiting Spot and Burstable Instances for Improving the Cost-efficacy of In-Memory Caches on the Public Cloud},'' in \emph{Proceedings of the Twelfth European Conference on Computer Systems}, 2017, p. 620–634.

\bibitem{Jin2021}
Z.~Jin, Y.~Zhu, J.~Zhu, D.~Yu, C.~Li, R.~Chen, I.~E. Akkus, and Y.~Xu, ``{Lessons learned from migrating complex stateful applications onto serverless platforms},'' in \emph{Proceedings of the 12th ACM SIGOPS Asia-Pacific Workshop on Systems}, 2021.

\bibitem{Fou2017}
S.~Fouladi, R.~S. Wahby, B.~Shacklett, K.~V. Balasubramaniam, W.~Zeng, R.~Bhalerao, A.~Sivaraman, G.~Porter, and K.~Winstein, ``{Encoding, fast and slow: low-latency video processing using thousands of tiny threads},'' in \emph{Proceedings of the 14th USENIX Conference on Networked Systems Design and Implementation}, 2017.

\bibitem{Gun2019}
J.~Gunasekaran, P.~Thinakaran, M.~Kandemir, B.~Urgaonkar, G.~Kesidis, and C.~Das, ``{Spock: Exploiting Serverless Functions for SLO and Cost Aware Resource Procurement in Public Cloud},'' in \emph{IEEE 12th International Conference on Cloud Computing (CLOUD)}, 2019.

\bibitem{Jai2020}
A.~Jain, A.~Baarzi, B.~Kesidis, G.~Urgaonkar, N.~Nader~Alfares, and M.~Kandemir, ``{SplitServe: Efficiently Splitting Apache Spark Jobs Across FaaS and IaaS},'' in \emph{Proceedings of the 21st International Middleware Conference}, 2020.

\bibitem{Zha2019}
C.~Zhang, M.~Yu, W.~Wang, and F.~Yan, ``{MArk: Exploiting cloud services for cost-effective, SLO-aware machine learning inference serving},'' in \emph{Proceedings of the 2019 USENIX Conference on Usenix Annual Technical Conference}, 2019.

\bibitem{Son2022}
M.~Son, S.~Mohanty, J.~Gunasekaran, A.~Jain, M.~Kandemir, G.~Kesidis, and B.~Urgaonkar, ``{Splice: An Automated Framework for Cost-and Performance-Aware Blending of Cloud Services},'' in \emph{22nd IEEE International Symposium on Cluster, Cloud and Internet Computing (CCGrid)}, 2022.

\bibitem{Son2023}
M.~Son, S.~Mohanty, J.~Gunasekaran, and M.~Kandemir, ``{MicroBlend: An Automated Service-Blending Framework for Microservice-Based Cloud Applications},'' in \emph{IEEE 16th International Conference on Cloud Computing (CLOUD)}, 2023.

\bibitem{Kho23}
A.~Khochare, T.~Khare, V.~Kulkarni, and Y.~Simmhan, ``{XFaaS: Cross-platform Orchestration of FaaS Workflows on Hybrid Clouds},'' in \emph{2023 IEEE/ACM 23rd International Symposium on Cluster, Cloud and Internet Computing (CCGrid)}, 2023.

\bibitem{Gor2024}
V.~A.~K. Gorantla, V.~Gude, S.~K. Sriramulugari, N.~Yuvaraj, and P.~Yadav, ``{Utilizing Hybrid Cloud Strategies to Enhance Data Storage and Security in E-Commerce Applications},'' in \emph{2024 2nd International Conference on Disruptive Technologies (ICDT)}, 2024.

\bibitem{terraform}
\BIBentryALTinterwordspacing
``{Terraform},'' {Accessed: 2025-07-31}. [Online]. Available: \url{https://developer.hashicorp.com/terraform}
\BIBentrySTDinterwordspacing

\bibitem{Tep2022}
H.~Teppan, L.~H. Flå, and M.~G. Jaatun, ``{A Survey on Infrastructure-as-Code Solutions for Cloud Development},'' in \emph{2022 IEEE International Conference on Cloud Computing Technology and Science (CloudCom)}, 2022.

\bibitem{cloud_native_observability}
\BIBentryALTinterwordspacing
{Lumigo}, ``{Cloud Native Observability: An Introduction \& 5 Tips for Success},'' {Accessed: 2025-07-31}. [Online]. Available: \url{https://lumigo.io/microservices-monitoring/cloud-native-observability-an-introduction-and-5-tips-for-success/}
\BIBentrySTDinterwordspacing

\bibitem{Fas2025}
U.~Faseeha, H.~Jamil~Syed, F.~Samad, S.~Zehra, and H.~Ahmed, ``{Observability in Microservices: An In-Depth Exploration of Frameworks, Challenges, and Deployment Paradigms},'' \emph{IEEE Access}, vol.~13, 2025.

\bibitem{github_gob}
\BIBentryALTinterwordspacing
``{Google Online Boutique: A Microservices Demo Application},'' {Accessed: 2025-07-31}. [Online]. Available: \url{https://github.com/GoogleCloudPlatform/microservices-demo}
\BIBentrySTDinterwordspacing

\bibitem{github_mediareview}
\BIBentryALTinterwordspacing
``{Spring Cloud Movie Recommendations},'' {Accessed: 2025-07-31}. [Online]. Available: \url{https://github.com/mdeket/spring-cloud-movie-recommendation}
\BIBentrySTDinterwordspacing

\bibitem{GCP}
``{Google Cloud Platform},'' \url{https://cloud.google.com/}.

\bibitem{serverless_nodejs_advantages}
\BIBentryALTinterwordspacing
A.~Das, ``{What Are the Advantages of Serverless Node.js Solutions?}'' {Accessed: 2025-07-31}. [Online]. Available: \url{https://dev.to/arunangshu_das/what-are-the-advantages-of-serverless-nodejs-solutions-4p7d}
\BIBentrySTDinterwordspacing

\bibitem{prometheus}
\BIBentryALTinterwordspacing
``{Prometheus},'' {Accessed: 2025-07-31}. [Online]. Available: \url{https://prometheus.io/}
\BIBentrySTDinterwordspacing

\bibitem{grafana}
\BIBentryALTinterwordspacing
``{Grafana},'' {Accessed: 2025-07-31}. [Online]. Available: \url{https://grafana.com/}
\BIBentrySTDinterwordspacing

\bibitem{locust}
\BIBentryALTinterwordspacing
``{Locust},'' {Accessed: 2025-07-31}. [Online]. Available: \url{https://locust.io//}
\BIBentrySTDinterwordspacing

\bibitem{Mal2020}
S.~Malla and K.~Christensen, ``{HPC in the cloud: Performance comparison of Function as a Service (FaaS) vs Infrastructure as a Service (IaaS)},'' \emph{Internet Technology Letters}, vol.~3, no.~1, Dec. 2019.

\bibitem{nodejs_runtime}
\BIBentryALTinterwordspacing
{Google Cloud}, ``{The Node.js runtime},'' {Accessed: 2025-07-31}. [Online]. Available: \url{https://cloud.google.com/run/docs/runtimes/nodejs}
\BIBentrySTDinterwordspacing

\bibitem{gke_enterprise_overview}
\BIBentryALTinterwordspacing
------, ``{GKE Enterprise Technical Overview},'' {Accessed: 2025-07-31}. [Online]. Available: \url{https://cloud.google.com/kubernetes-engine/enterprise/docs/concepts/overview}
\BIBentrySTDinterwordspacing

\bibitem{Ust2023}
T.~Schirmer, N.~Japke, S.~Greten, T.~Pfandzelter, and D.~Bermbach, ``{The Night Shift: Understanding Performance Variability of Cloud Serverless Platforms},'' in \emph{Proceedings of the 1st Workshop on Serverless Systems, Applications and MEthodologies}, ser. SESAME '23, 2023.

\bibitem{Gol2024}
M.~Golec, G.~K. Walia, M.~Kumar, F.~Cuadrado, S.~S. Gill, and S.~Uhlig, ``{Cold Start Latency in Serverless Computing: A Systematic Review, Taxonomy, and Future Directions},'' \emph{ACM Computing Surveys}, vol.~57, no.~3, Nov. 2024.

\bibitem{cpu_boost_GCF}
\BIBentryALTinterwordspacing
{Google Cloud}, ``{Announcing Startup CPU Boost for Cloud Run and Cloud Functions},'' accessed: 2025-06-15. [Online]. Available: \url{https://cloud.google.com/blog/products/serverless/announcing-startup-cpu-boost-for-cloud-run--cloud-functions}
\BIBentrySTDinterwordspacing

\bibitem{GithubActions}
\BIBentryALTinterwordspacing
``{Github Actions},'' {Accessed: 2025-07-31}. [Online]. Available: \url{https://github.com/features/actions}
\BIBentrySTDinterwordspacing

\end{thebibliography}

\end{document}